\documentclass[%
 reprint,
 amsmath,amssymb,
 aps,
]{revtex4-2}

\usepackage{xcolor}
\usepackage{graphicx}
\usepackage{dcolumn}
\usepackage{bm}
\usepackage{amsmath}
\usepackage{subcaption}
\usepackage[justification=raggedright,singlelinecheck=false]{caption}
\usepackage{graphicx}
\usepackage[version=4]{mhchem}

\newcounter{pr}

\definecolor{darkred}{rgb}{.8,0,0}

\begin{document}

\title{
Critical Charge and Current Fluctuations across a Voltage-Driven Phase Transition
}


\author{José F. B. Afonso}
\email{  josefafonso@tecnico.ulisboa.pt}
\affiliation{CeFEMA-LaPMET, Physics Department, Instituto Superior T\'ecnico, Universidade de Lisboa Av. Rovisco Pais, 1049-001 Lisboa, Portugal}

\author{Stefan Kirchner}
\email{kirchner@nycu.edu.tw}
\affiliation{Department of Electrophysics, National Yang Ming Chiao Tung University, Hsinchu 30010, Taiwan}

\author{Pedro Ribeiro}
\email{ribeiro.pedro@tecnico.ulisboa.pt}
\affiliation{CeFEMA-LaPMET, Physics Department, Instituto Superior T\'ecnico, Universidade de Lisboa Av. Rovisco Pais, 1049-001 Lisboa, Portugal}
\affiliation{Beijing Computational Science Research Center, Beijing 100193, China}

\begin{abstract}
We investigate bias-driven non-equilibrium quantum phase transitions in a paradigmatic quantum-transport setup: an interacting quantum dot coupled to non-interacting metallic leads. 
Using the Random Phase Approximation, which is exact in the limit of a large number of dot levels, we map out the zero-temperature non-equilibrium phase diagram as a function of interaction strength and applied bias. We focus our analysis on the behavior of the charge susceptibility and the current noise in the vicinity of the transition.
Remarkably, despite the intrinsically non-equilibrium nature of the steady state, critical charge fluctuations admit an effective-temperature description, $T_{\mathrm{eff}}(T,V)$, that collapses the steady-state behavior onto its equilibrium form. In sharp contrast, current fluctuations exhibit genuinely non-equilibrium features: the fluctuation–dissipation ratio becomes negative in the ordered phase, corresponding to a negative effective temperature for the current degrees of freedom.
These results establish current noise as a sensitive probe of critical fluctuations at non-equilibrium quantum phase transitions and open new directions for exploring voltage-driven critical phenomena in quantum transport systems.

    \vspace{0.5 cm}
    
  \textbf{Keywords:} electron transport, quantum dot, quantum phase transitions, non-equilibrium dynamics
\end{abstract}

\maketitle

\section{Introduction}
\label{sec:Introduction}

Non-equilibrium quantum systems constitute a rich and still incompletely explored frontier of condensed-matter physics. Away from equilibrium, dynamics are no longer constrained by fluctuation--dissipation relations, enabling behavior that has no thermal counterpart. Driven open systems may therefore exhibit steady-state phase transitions whose universality classes differ from those dictated by equilibrium statistical mechanics. Yet, despite this apparent freedom, several non-equilibrium critical phenomena appear to admit an effective-temperature description, suggesting that—although true thermalization is absent—critical behavior may resemble its equilibrium counterpart when expressed in terms of a suitably defined effective temperature $T_{\mathrm{eff}}$. Determining when non-equilibrium systems deviate from, or instead mimic, equilibrium criticality remains a central problem in the theory of driven quantum matter.

In this work, we study charge and current fluctuations in a paradigmatic model of non-equilibrium quantum transport: an interacting, particle--hole--symmetric quantum dot tunnel-coupled to metallic leads and driven by a finite bias voltage. Using the Random Phase Approximation (RPA), which becomes exact in the limit of a large number of dot levels, we determine the zero-temperature non-equilibrium phase diagram and analyze the associated critical fluctuations.

We find that charge fluctuations admit an effective-temperature description: when expressed in terms of $T_{\mathrm{eff}}$, the charge susceptibility collapses onto its equilibrium critical form. In contrast, current fluctuations display genuinely non-equilibrium behavior. The current noise diverges more strongly than the charge susceptibility at the transition, and the corresponding fluctuation--dissipation ratio becomes negative in the ordered phase, implying a negative effective temperature for the current degrees of freedom.

Non-equilibrium steady-state phase transitions in driven open quantum systems have been investigated in a wide range of settings \cite{Maghrebi2016_NoneqSteadyKeldysh,Sieberer2016_KeldyshReview,Zhang2021_KerrTransition,DiVentra2008,Breuer2002}. These include non-perturbative field-theoretical formulations \cite{Podolsky2024_SKnonperturbative,Cavina2023_ConvenientContour}, driven-dissipative lattice models and non-equilibrium Mott transitions \cite{Sankar2018_MottKeldysh,Zhang2021_NoneqQCSteadyState}, cavity-QED realizations of Dicke-type criticality \cite{Torre2013_KeldyshDicke,Brennecke2013_DickeFluctuations}, and driven Bose--Einstein condensates \cite{Szymanska2007,Szymanska2006}. In many of these systems, the breaking of detailed balance leads to emergent dynamical universality classes \cite{Kamenev2011,Sieberer2016_KeldyshReview} that differ qualitatively from those governing equilibrium quantum critical points \cite{SachdevQPT,VojtaQPT,ColemanSchofield,RevModPhys.92.011002,Stewart2001}.

At the same time, a number of non-equilibrium critical points exhibit effective-temperature behavior, whereby correlation and response functions can be organized as if they obeyed an equilibrium fluctuation--dissipation relation characterized by $T_{\mathrm{eff}}$ \cite{RibeiroSiKirchner2013,Ribeiro2015,Zamani.16}. Clarifying when such descriptions apply, and when genuinely new universality classes emerge, remains an open challenge.

The concept of an effective temperature has therefore attracted sustained attention in driven critical systems \cite{HohenbergHalperin,CugliandoloKurchanPeliti,CalabreseGambassi,CugliandoloReview}. In classical systems, $T_{\mathrm{eff}}$ can characterize relaxational dynamics near criticality \cite{CalabreseGambassi}, although its interpretation becomes subtle at fully interacting fixed points \cite{CugliandoloReview}. In quantum systems, non-equilibrium driving induces decoherence processes set by bias or external drive. Remarkably, scaling regimes such as $\omega/T$ or $\omega/V$ have been identified near quantum criticality even far from equilibrium, consistent with interacting fixed points \cite{MitraMillis2006,KirchnerSi2009,RibeiroSiKirchner2013}. In such cases, the fluctuation--dissipation ratio may define a fixed-point--specific $T_{\mathrm{eff}}$, allowing steady-state correlation and response functions to collapse onto equilibrium scaling forms upon replacing $T$ by $T_{\mathrm{eff}}$ \cite{Ribeiro2015}. Related effective-equilibrium scenarios have also been reported in resistive switching \cite{Han2018_NoneqMF}. Notably, this description can extend beyond two-point susceptibilities to higher-order correlation functions \cite{KirchnerSi2009,RibeiroSiKirchner2013}.

Quantum dots provide a particularly well-controlled platform for exploring these issues. Modern nanofabrication techniques enable tunable few-electron devices with precise control over interactions, tunnel couplings, and non-equilibrium driving \cite{LPKouwenhoven_2001,Datta2005,Nazarov2009}. Such systems offer direct experimental access to observables sensitive to critical fluctuations, including the electric current, its noise, and the dot charge susceptibility \cite{Buttiker2000}. These quantities can be measured with high precision, making quantum dots an ideal setting for probing non-equilibrium universality.

This includes higher-point correlators such as current noise, which has recently attracted renewed interest in the context of beyond-Landau criticality \cite{Chen2023}. Current--current correlation functions encode the temporal structure of transport fluctuations and provide a direct probe of deviations from equilibrium statistics. Near quantum phase transitions, where critical fluctuations become long-ranged, such correlations strongly renormalize transport properties and may destabilize mean-field steady states. An effective low-energy description beyond the saddle-point approximation is therefore required.

Here, we focus on a minimal quantum-dot model capturing the essential ingredients of non-equilibrium transport and critical behavior. We consider an effective attractive on-site interaction, corresponding to a negative Coulomb matrix element. Although negative-$U$ interactions are not generic, they arise in experimentally motivated settings such as valence-skipping systems \cite{Matsuura.22} or through coupling to local phonon modes \cite{FirsovLang.63}. In this regime, low-energy dynamics is dominated by charge fluctuations, enabling a controlled description in terms of charge degrees of freedom alone. Despite its simplicity, the model retains all key elements—local interactions, coherent tunneling to biased leads, and steady-state driving—providing a suitable framework for analyzing electronic transport, current fluctuations, and current--current correlations at a non-equilibrium phase transition.


\begin{figure*}[htbp]
    \centering
    \includegraphics[width=\textwidth]{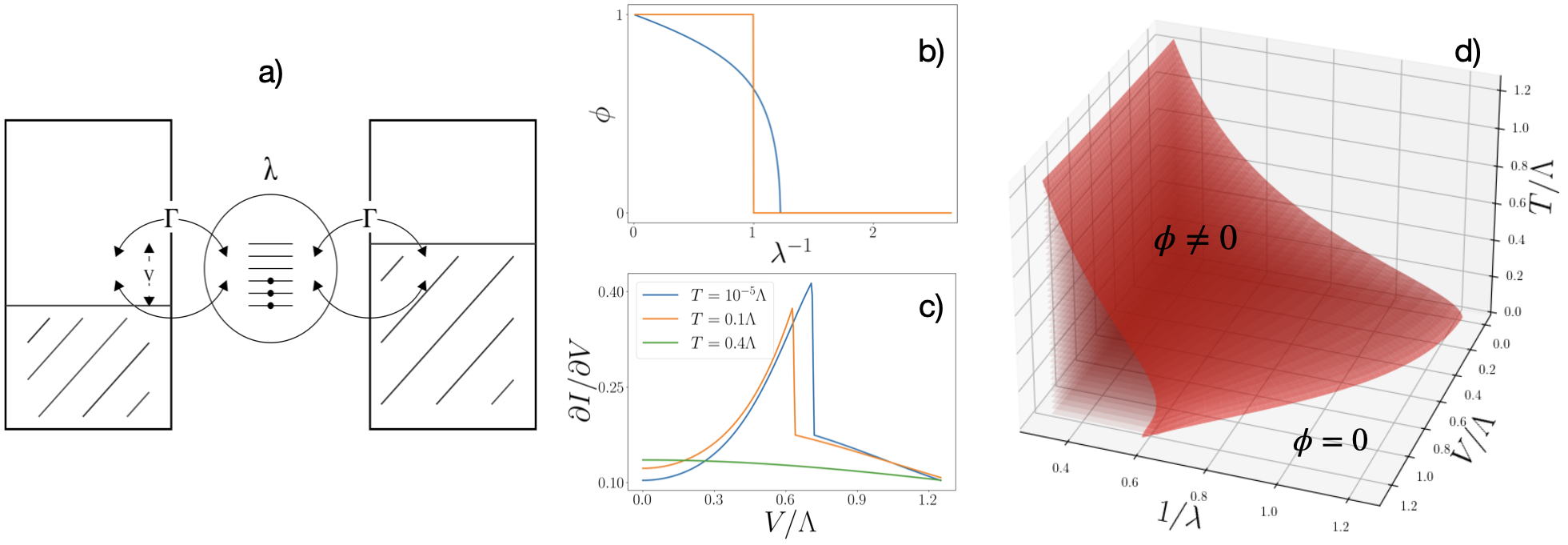}
    \caption{
    (a) Schematic representation of the quantum transport setup.
    (b) Order parameter $\phi$ as a function of the inverse interaction strength $\lambda^{-1}$ for both zero $(T=0)$ and finite $(T>0)$ temperature.
    (c) Conductance $G=dI/dV$ as a function of voltage $V$ for several different temperatures.
    (d) Non-equilibrium phase diagram as a function of interaction strength $\lambda^{-1}$, temperature $T$, and bias voltage $V$, showing the continuous transition between the ordered and disordered phases.
    }
    \label{fig:MeanField}
\end{figure*}

\section{Model and Methods}
\label{sec:Model}


To explore charge noise near criticality, we employ a paradigmatic, tractable model featuring transport through an interacting quantum dot,  which has the added advantage of being exactly solvable in the thermodynamic limit. 

The system, illustrated in Fig.~\ref{fig:MeanField}(a), is comprised of an interacting quantum dot connected to metallic leads. The total Hamiltonian is given by  
\begin{equation}
    H = H_\text{dot} +  H_\text{dot-leads}  +  H_\text{leads}, \label{eq:Model:H}
\end{equation}
here 
\begin{equation} 
    H_\text{dot} = \sum_{\alpha=1}^M \varepsilon_{\alpha} d_{\alpha}^{\dagger} d_{\alpha} -  \frac{\lambda}{M} \left( \sum_{\alpha=1}^M d_{\alpha}^{\dagger} d_{\alpha} - \frac{M}{2}\right)^2 
\end{equation} 
is the dot Hamiltonian with $\varepsilon_{\alpha=1,...,M}$ the individual energies of the dot's $M$ levels and $\lambda$ a negative charging energy around the half-filled condition.
The coupling to the leads is given by 
\begin{equation} 
    H_\text{dot-leads} = \frac{1}{\sqrt{V}} \sum_{l, \nu, k, \alpha}  \left(c_{l \nu k}^{\dagger} \tau^{(l)}_{\nu,\alpha} d_{\alpha} + d_{\alpha}^{\dagger} \tau^{(l)\dagger}_{\alpha,\nu} c_{l \nu k}\right)
\end{equation}
where $l=\text{L},\text{R}$ labels the left or right lead, $\nu$ the transport channels of the leads and $k$ the momentum of the electron. The Hamiltonian of the leads is given by  $H_\text{leads} = \sum_{l,\nu,k} \Omega_{l,\nu,k} c_{l\nu k}^{\dagger} c_{l\nu k}$.
As we consider metallic wide band leads, the details of the dispersion relation $\Omega_{l,\nu,k}$ are unimportant \cite{Ribeiro2016}.

The lead $l$, taken to be infinite, is assumed to have been prepared in equilibrium, characterized by chemical potential $\mu_l$ and temperature $T_l$, before the coupling to the dot was turned on. In the following, we address the properties of the steady-state, where the transient regime has vanished. 

When $\sum_\alpha \varepsilon_{\alpha}= 0$ the Hamiltonian of the isolated dot possesses particle-hole symmetry with respect to its chemical potential $\mu=0$. However, for large enough $\lambda$ the ground-state spontaneously breaks this symmetry.  
Since the number of particles on the dot commutes with the Hamiltonian, this transition is first order. 
In the presence of the leads, particle-hole symmetry is maintained for $\mu_\text{L}=\mu_\text{R}=V/2$ and symmetric coupling even in the presence of a finite bias ($V$). As shown in Fig.~\ref{fig:MeanField}-(b), any finite dot-lead hybridization renders the symmetry breaking transition continuous.  

It is convenient to take the energy levels of the dot to obey $\varepsilon_\alpha = \Lambda (-1 + (2\alpha-1)/M) $, which ensues a constant density of states  within the energy cutoff $\Lambda$. Nonetheless the results are qualitatively the same for other choices as long as particle-hole symmetry is preserved. We also take $\tau^{(l)}_{\nu,\alpha} = \tau \delta_{\nu,\alpha}$, i.e. each dot level tunnels to a transport channel with a different symmetry. Therefore, in the wide-band limit, the effects of the leads on the dot enter only through  the hybridization matrix, $\boldsymbol{\Gamma}_{\alpha,\alpha'} = \Gamma \delta_{\alpha,\alpha'}$, where $\Gamma \propto |\tau|^2 \rho(0) $, with $\rho(0)$ denoting the density of states of the leads.

In the following, we analyze the non-equilibrium phase diagram of the model, focusing on the charge fluctuations on the dot, the current and its fluctuations.
This will be done using an action functional approach on the Keldysh contour.

\subsection{Keldysh Action}
\label{sec:KeldyshAction}

To characterize the charge dynamics we consider the response $\chi^\text{R}(t,t') = -i \left\langle \left[ N_d(t) , N_d(t') \right]\right\rangle $ and fluctuation functions $\chi^\text{K}(t,t') = -i \left\langle \left\{ N_d(t) , N_d(t') \right\}\right\rangle $. $N_d = \sum_\alpha d_\alpha^\dagger d_\alpha $ is the total number of particles on the dot, and similarly, we define the number of particles in lead $l$ as $N_l = \sum_ {\nu k} c_{l \nu k}^\dagger c_{l \nu k}$. We also study the current, which, with the convention $J_{l} = d N_l/dt = i [H, N_l] $, is explicitly given by 
\begin{equation}
    J_l  =  \frac{i}{\sqrt{V}} \sum_{ \nu, k, \alpha}  \left(c_{l \nu k}^{\dagger} \tau^{(l)}_{\nu,\alpha} d_{\alpha} - d_{\alpha}^{\dagger} \tau^{(l)\dagger}_{\alpha,\nu} c_{l \nu k}\right), 
    \label{eq:KeldyshAction:CurrentOperatorKeldysh}
\end{equation}
and its response $S^\text{R}_{l,l'}(t,t') = -i \left\langle \left[ J_l(t) , J_{l'}(t') \right]\right\rangle $ and fluctuation  $S^\text{K}_{l,l'}(t,t') = -i \left\langle \left\{ J_l(t) , J_{l'}(t') \right\}\right\rangle $ dynamics. Specifically, we set $l=l'=\text{L}$ in what follows. 

It is worth noting the differences between the current operator in Eq.~(\ref{eq:KeldyshAction:CurrentOperatorKeldysh}) and the one typically considered in a \textit{Landauer-Buttiker} approach. The current operator as defined above considers the measurement to occur near the lead in contrast with the Landauer-Buttiker formalism for which the current is measured inside the lead at a large distance from the dot. Although the average value of the current should coincide in both cases, there are differences in current fluctuations. 

We obtain the expressions for these observables, using a functional  approach within the RPA approximation. To do this, we consider the generating function
\begin{equation}
    Z[j,u] = \int Dc \int Dd ~e^{i S[j,u]}
    \label{eq:KeldyshAction:Z}
\end{equation}
where $S$, explicitly given by 
\begin{align}
    S[j,u] & =  \int_C dz \left[  c^* g_c^{-1} c + d^* g_d^{-1} d   + \frac{ \lambda}{M} \left( N_d(z) - \frac{M}{2}\right)^2 \right. \nonumber \\
    &- H_\text{dot-leads}(z)  +   j_l(z) J_l(z) + u(z) N_d(z)  \Bigg]
    \label{eq:KeldyshAction:Z}
\end{align}
is the action on the Keldysh contour where, $g_c$ and $g_d$ are the bare Green's functions of the leads and dot, respectively, and $j(z)$ and $u(z)$ are source fields conjugate to current and charge of the dot. (The convention of summing over repeated indices has been adopted.) 

Decoupling the interacting term via the introduction of a bosonic  Hubbard-Stratonovich decoupling field  $\phi$  results in an effective action that is  quadratic in the Grassman fields. Integrating out the Grassmann fields results in a bosonic action given by 
\begin{align}
    S[\phi,j,u] &= -M \int_C dz ~  \lambda \left[ \phi^2(z) + \phi(z) \right] \nonumber \\
   &  -i  \mathrm{Tr}  \ln{ \left[ i \left( G_0^{-1} - \tilde\Sigma[j] + 2 \lambda \phi - u \right) \right]}.
    \label{eq:KeldyshAction:GeneralActionPhi}
\end{align}
where 
\begin{gather}  
    G_0^{-1}=  g_d^{-1} - \Sigma_0, \\
    \Sigma_0 = \frac{1}{V} \tau^\dagger g_0  \tau \\ 
    \tilde{\Sigma}[j] = \frac{1}{V} \tau^\dagger \left( -i g_0 j + i j g_0 + j g_0 j \right) \tau.
\end{gather}
In the following we perform a Keldysh rotation into classical and quantum variables where $\phi$ acquires a two component structure $\boldsymbol \phi = (\phi^{cl}, \phi^{q})^T $ and it is useful to define the matrix $\hat{\phi} = \phi^{cl} + \phi^q \sigma_1$ acting in Keldysh space, with $\sigma_i~(i=1,2,3)$ are the Pauli Matrices.

\subsection{Fluctuation Dissipation Ratios and Effective Temperature}
\label{sec:FDR}
The fluctuation-dissipation theorem links the response of a system in thermal equilibrium to its fluctuation spectrum which, in terms of Schwinger-Keldysh Green's functions, can be expressed as $\text{Q}_\text{FDR}(\omega) = \tanh (\beta \omega/2)$, which becomes a step function at $T=0$ for correlators of bosonic fields. For nonthermal steady states, this generalizes to the
fluctuation dissipation ratio (FDR)  $\text{Q}_{\text{FDR}}^X = \Im \chi_{X }^R/\Im \chi_{X}^K$, with $\chi_X$ a correlator of observable $X$ (e.g. charge: $X=N_d$), which, away from equilibrium, may not only depend on $T$ and $\omega$ and the statistics of $\chi_X$. The FDR  of $\chi_X$ thus contains information about the linear-response regime, i.e., the regime where the system appears to be thermal.
It also allows for a definition of an effective temperature $T_{\text{eff}}^X $ in the nonlinear regime.  

Interestingly, as we show in Appendix~\ref{sec:Thermometry}, $\text{Q}_{\text{FDR}}^X(\Omega)$   can be measured by monitoring the occupation of an harmonic oscillator with frequency $\Omega$ weakly coupled to that observable, i.e. $H_\text{HO} = \Omega a^\dagger a + \eta X (a+a^\dagger ) $. In the $\eta\to 0$ limit, we obtain (see Appendix~\ref{sec:Thermometry}):   
\begin{gather}
\left\langle a^{\dagger}a \right\rangle =\frac{1}{2}\left[Q_{\text{FDR}}^{X}\left(\Omega\right)-1\right].
\end{gather}

The non-equilibrium FDR motivates the definition of an effective temperature 
\begin{gather}
T_\text{eff}^X = \lim_{\omega\to0}\frac{1}{2} [\partial_\omega \text{Q}^X_\text{FDR}(\omega)]^{-1},
\label{eq:T_eff}
\end{gather}
which can be seen as the temperature inferred from the occupation of a harmonic oscillator with frequency $\omega\to0$, weakly coupled to the $X$ degrees of freedom of the dot.

\subsection{Mean Field}
\label{sec:MeanField}
The mean-field equations are obtained by minimizing the action of Eq.~\ref{eq:KeldyshAction:GeneralActionPhi} with respect to $\phi$, setting the sources $u$ and $j$ to zero. As the action globally scales with the number of levels on the dot, $M$, this saddle-point approximation becomes exact in the limit of large $M$ ($M\rightarrow \infty$).

The self-consistent mean-field equations in the (time-translationally invariant) steady-state yield
\begin{gather}
    \phi^{cl}_0 = -\frac{i}{2M} \int \frac{d\omega}{2\pi} ~\text{tr}\left[\tilde{G}^K_0(\omega)\right], 
\end{gather}
and $\phi^q_0 = 0$ where $\tilde G^K_0(\omega)$ is the Mean-Field Keldysh Green's function in the frequency domain and trace is performed over the dot's energy levels.

It is useful to introduce the mean-field single particle non-hermitian operator, $\mathbf{K} = H_\text{MF} -i\Gamma$, where $H_\text{MF} = \sum_{\alpha} d^{\dagger}_{\alpha} (\varepsilon_{\alpha} - 2\lambda \phi^{cl}_0) d_{\alpha}$ and $\Gamma$ represents the coupling value to the leads. We assume the coupling of the left and right leads are the same, so $\Gamma = (\Gamma_L + \Gamma_R)/2 \equiv \Gamma$.
With this notation, the mean-field Green's functions are given by 
\begin{gather}
\tilde{G}^R_0(\omega) = \left[ \tilde{G}^A_0(\omega)\right]^{\dagger} = \left( \omega - \mathbf{K} \right)^{-1} =  \sum_{\alpha} | \alpha \rangle \frac{1}{\omega - \lambda_{\alpha}} \langle \tilde{\alpha} |, \\
\tilde{G}^K_0(\omega) = -i\Gamma \sum_{l \alpha} | \alpha \rangle \frac{ F_l(\omega)}{(\omega - \lambda_\alpha) (\omega - \lambda_\alpha^*) } \langle \alpha |,
\end{gather}
where the labels $R$, $A$ and $K$ refer respectively to the \textit{retarded}, \textit{advanced} and \textit{Keldysh Green's} Functions. The $|\alpha \rangle$ and $\langle \tilde{\alpha} |$ are the right and left eigenvectors of the $\mathbf{K}$ operator, with eigenvalue $\lambda_\alpha$ 
Here, $\mathbf{K}$ basis is diagonal with  $\lambda_{\alpha} = \varepsilon_{\alpha} - 2\lambda \phi^{cl}_0 -i \Gamma$.  $F_l(\omega) = \tanh{ \beta/2 (\omega-\mu_l)}$ encodes temperature and chemical potential of lead $l$.  
In what follows, the leads are held at the same temperature, i.e. $T_\text{R}=T_\text{L}=T$. 

The non-equilibrium phase diagram as a function of the interaction strength, $\lambda$, temperature $T$, and  bias voltage $V$ is depicted in Fig.~\ref{fig:MeanField}-(d).
As already mentioned, we take the energy distribution of a dot to be constant, ranging from $-\Lambda/2$ to $\Lambda/2$, where $\Lambda = 2$ but confirmed that a different density of states respecting particle-hole symmetry gives the same qualitative results. The ordered phase that exists in equilibrium for a sufficiently large coupling constant can be destroyed in a continuous fashion either by increasing the temperature or the bias voltage. 
The conductance $G=d I/dV$ for different temperatures is given in Fig.~\ref{fig:MeanField}-(c). Within the ordered phaese upon increasing voltage $G$ increases and passes by a discontinuity at the transition, decreasing with voltage in the disordered phase.

\begin{figure*}[htbp]
    \includegraphics[width=\textwidth]{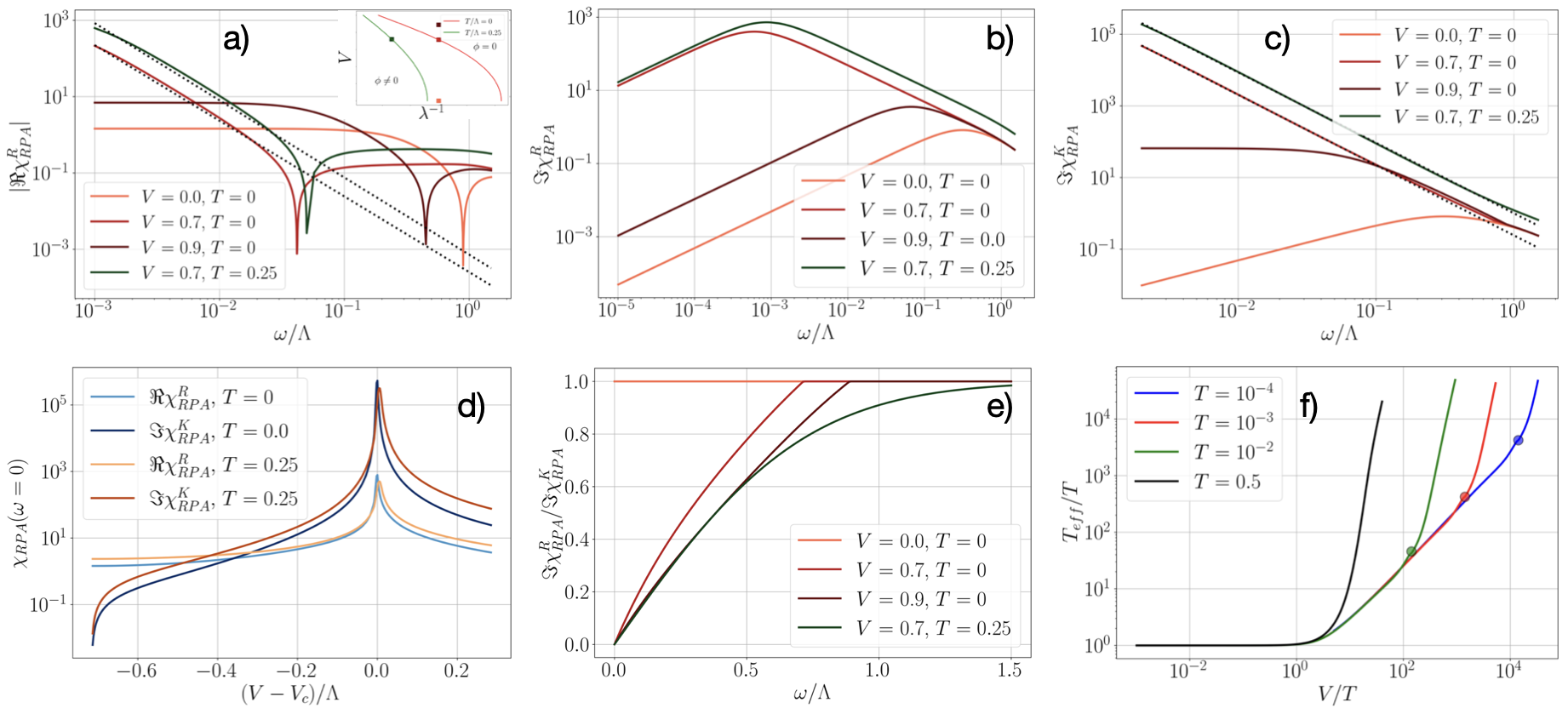}
    \caption{Charge Susceptibility across the non-equilibrium phase transition. (a-c) Real part (a), imaginary part (b) of the retarded susceptibility $\chi_{\rm RPA}^R$, and the Keldysh component (c) $\chi_{\rm RPA}^K$ as a function of frequency $\omega$. Curves are shown for $T=0$ at $V<V_c$, $V=V_c$, and $V>V_c$ (orange scale), and for $T>0$ at $V=V_c (T)$. (d) Zero-frequency retarded $\chi^{\rm RPA}_R(\omega=0)$ and Keldysh $\chi^{\rm RPA}_K(\omega=0)$ components as a function of voltage, showing the critical divergence at $V_c$. (e) Fluctuation-dissipation ratio (FDR) as a function of frequency $\omega$, using the same parameters and color-coding as (a-c). (f) Effective temperature $T_{\rm eff}/T$, derived from the zero-frequency FDR, plotted as a function of $V/T$ for several lead temperatures $T$. }
    \label{fig:Susceptibility}
\end{figure*}

\begin{figure*}[htbp]
    \includegraphics[width=1.0\textwidth]{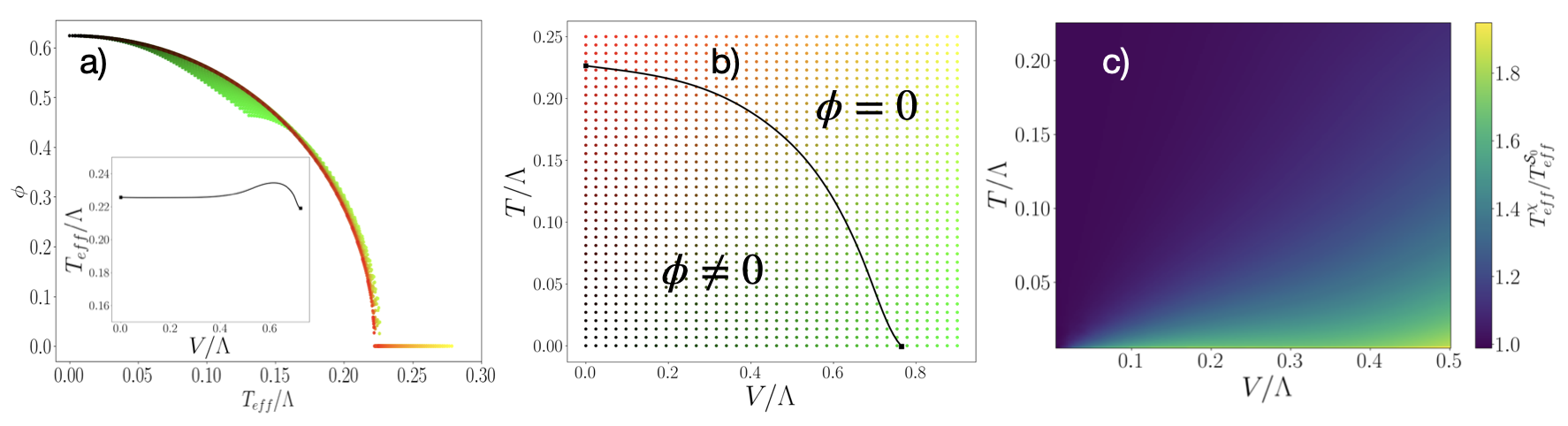}
    \caption{
    (a) Order parameter $\phi$ as a function of the effective temperature $T^{\chi}_{\rm eff}$. (b) The $V–T$ phase diagram. The color-coded points indicate the $(V,T)$ values plotted in (a). The inset shows the effective temperature along the critical line $V_c(T)$.
    (c) Ratio of effective temperatures derived from current noise $T^{\mathcal{S}_0}_{\rm eff}$ and charge susceptibility $T^{\chi}_{\rm eff}$ in the non-interacting limit, shown as a function of voltage $V$ and temperature $T$.
    }
    \label{fig:phy_vs_teff_ShotVT_non_interacting}

\end{figure*}

\section{Charge Susceptibility}
\label{sec:RPA}

To obtain the charge susceptibility within RPA, we consider the fluctuation of the order parameter, $\delta \vec{\phi}^T = \vec{\phi}^T - \vec{\phi}_0^T$, around the saddle point value, $\vec{\phi}_0$ and neglect the the current source, i.e. $j=0$. 
In this approximation the action becomes 
\begin{align}
    S_\text{RPA}[\phi, u] = \int dt ~ 
    \delta \vec{\phi}^T (2\lambda \chi_0) \vec{u}  +\vec{u}^T (2\lambda \chi_0) \delta\vec{\phi} \notag \\
    - \delta \vec{\phi}^T D^{-1} \delta \vec{\phi} 
    - \vec{u}^T \chi_0 \vec{u},
\end{align} 
where
\begin{align}
    D^{-1} = \lambda( \sigma_1 + 4\lambda \chi_0),
\end{align}
is the inverse of the $\phi$ propagator, and the bare "bubble" is given by
\begin{align}
    \chi_0^{\alpha \beta} (t, t') = - \frac{i}{2} \mathrm{Tr} \left[ \tilde{G}_0(t, t') \gamma^{\alpha} \tilde{G}_0(t',t) \gamma^{\beta} \right].
    \label{eq:RPA:PiMatrix}
\end{align}
where $\alpha$ and $\beta$ run over the \textit{classical (cl)} and \textit{quantum (q)} components and 
 $\gamma^{cl} = I$ and $\gamma^q = \sigma_1$ are $2\times2$ matrices in the \textit{Keldysh} space.  

Integrating out the bosons and varying with respect to the sources, we get
\begin{equation}
   \chi_\text{RPA}(z,z') 
   = i \partial_{u(z)} \partial_{u(z')} \ln Z_\text{RPA}  
\end{equation}
with $\chi_\text{RPA} =  \chi_0 \left( 1 + 4\lambda \chi_0\right)^{-1}  $, or in Keldysh space
\begin{gather}
    \pmb{\chi}_\text{RPA} = \begin{pmatrix} 0 & \chi^A_\text{RPA} \\ \chi^R_\text{RPA} & \chi^K_\text{RPA} \end{pmatrix} = \notag \\
    = \begin{pmatrix}
        0                                          & \chi^A_0 \left( 1 + 4\lambda \chi^A_0\right)^{-1}                                     \\
       \chi^R_0 \left( 1 + 4\lambda \chi^R_0\right)^{-1}  & \left( 1 + 4\lambda \chi^R_0\right)^{-1}  \chi_0^K \left( 1 + 4\lambda \chi^R_0\right)^{-1} \end{pmatrix}.
        \label{eq:RPA:PiRPA}
\end{gather}



The results for the charge susceptibility are shown in Fig.~\ref{fig:Susceptibility}.
Panels (a)-(c) show respectively the real and imaginary parts of the retarded susceptibility and the Keldysh component for several values of the voltage,  at $V<V_c$, $V=V_c$ and $V>V_c$, both for $T=0$ and $V=V_c$ at $T=0.25 \Lambda$. For each temperature, the values of $\lambda$, {\itshape i.e.}, $\lambda/\Lambda = 0.5 $ and $\lambda/\Lambda = 0.6 $, were chosen to give the same critical voltage $V_c= 0.7  \Lambda$  (see inset of (a) and color code therein). 

In equilibrium $\chi_\text{RPA}^R(\omega=0) \sim |T-T_c|^{-1}$.

Away from equilibrium, we find $\chi_\text{RPA}^R(\omega=0) \sim |t|^{-1}$ with $t = (T_\text{eff}(T,V)-T_c)/T_c$, in particular the voltage behavior obeys $\chi_\text{RPA}^R(\omega=0)\sim |V-V_c|^{-1}$, as shown in Fig.~\ref{fig:Susceptibility}-(d).  


The scaling behavior of the retarded susceptibility follows from the bare expression that, for low-frequency, can be parameterized by
\begin{gather}
\chi_{0}^{R}\left(\omega\right)\simeq-\left[\left(\frac{1}{4\lambda}-at\right)+b\omega^{2}-ic\,\omega\right],
\end{gather}
with $a,b$ and $c$ are positive constants and $t = (T_\text{eff}(T,V)-T_c)/T_c$.   
The retarded RPA susceptibility, given in Eq.~\eqref{eq:RPA:PiRPA}, thus yield
\begin{gather}
\chi_{\text{RPA}}^{R}\left(\omega\right)=\chi_{\text{RPA, reg}}^{R}\left(\omega\right)+\chi_{\text{RPA,ireg}}^{R}\left(\omega\right),
\end{gather}
where 
\begin{gather}
\chi_{\text{RPA, reg}}^{R}\left(\omega\right)=\frac{1}{4\lambda}+\frac{b}{16c\lambda^{2}(c+ib\omega)}\\
\chi_{\text{RPA, ireg}}^{R}\left(\omega\right)=\frac{\frac{1}{-at+b\omega^{2}-ic\omega}-\frac{b}{c(c+ib\omega)}}{16\lambda^{2}}
\end{gather}
are respectively the regular and irregular components. 
For $\omega=0$ we obtained the divergence of the static susceptibility with the effective reduced temperature $\chi_{\text{RPA, ireg}}^{R}\left(\omega=0\right) \simeq-\frac{1}{16a\lambda^{2}t}$ which extends the equilibrium result. 
At $t=0$, the irregular part diverges as $\lim_{t\to0}\chi_{\text{RPA, ireg}}^{R}\left(\omega\right)=\frac{i}{16c\lambda^{2}\omega}$ with decreasing frequency. 

Assuming a finite effective temperature (see Fig.~\ref{fig:Susceptibility}-(f)) , it follows that the correlation function $\chi_\text{RPA}^K$, remains finite away from criticality but diverges for $\omega \rightarrow 0$, as $|\omega|^{-2}$, at the phase transition.

Fig.~\ref{fig:Susceptibility}-(e) depicts the fluctuation dissipation ratios $\text{Q}_\text{FDR}^{N_d} = \Im \chi_\text{RPA}^R/\Im \chi_\text{RPA}^K$ as a function of frequency for the cases of panels (a)-(c).  
For any non-vanishing voltage, the derivative of $\text{Q}_\text{FDR}^{N_d}$ at $\omega =0$ is finite.  
$\text{Q}_\text{FDR}^{N_d}$ interpolates between a strongly renormalized non-equilibrium value at small frequencies to its equilibrium form at the temperature of the leads for large frequencies.  
At zero temperature, this passage arises at a non-analytical point for $\omega=V$. For finite temperature, this becomes a crossover scale at $\omega\sim V$. 

Finally, Fig.~\ref{fig:Susceptibility}-(f) depicts the effective temperature (Eq.~\ref{eq:T_eff}) over the temperature of the leads as a function of $V/T$ for several values of $T$.  $T_\text{eff}$ interpolates between $T$ in equilibrium and a $\sim V^4$ behavior at large voltages. $T_\text{eff}$ is non-analytic at $V=V_c(T)$ which can {\itshape e.g.} be inferred from  panel (d).
There are three qualitative regimes, a linear response regime $V<T$, where $T_\text{eff} \simeq T$; a non-linear response region $V>T, T_\text{eff}<T_\text{C}$; and the large bias limit $V\gg T, T>T_\text{C}$.   

Interestingly, for the charge susceptibility we can show (see Appendix \ref{sec:FDR_RPA}) that the FDR computed within RPA is the same as the bare one, i.e.  
\begin{gather}
\frac{\Im \chi^R_\text{RPA}}{\Im \chi^K_\text{RPA}} = \frac{\Im \chi^R_0}{\Im \chi^K_0}. 
\end{gather}
This explains that $\text{Q}_\text{FDR}^{N_d}$ becomes non-analytic at the transition through dependence of $\chi_0$ of the order parameter. 
This argument also shows that, at the RPA level, the effective temperature of the order parameter does not diverge  at a non-equilibrium critical point even if its susceptibility diverges.  

Fig.~\ref{fig:phy_vs_teff_ShotVT_non_interacting}-(a) displays the order parameter $\phi$ as a function of the effective temperature $T_\text{eff}$ for several representative points in the $V$–$T$ phase diagram (see color code in (b)).
Strikingly, the data collapse onto a single curve near the transition, indicating that the behavior of $\phi$ in this regime may admit an equilibrium-like description when expressed in terms of $T_\text{eff}$ rather than the actual temperature $T$.


\section{Current Noise}
\label{sec:CurrentNoise}

We now turn to the behavior of the current and its fluctuations across the non-equilibrium critical point.  
For this, we include the current source term $j$ in the action and compute the average current and current noise   
via differentiation of the RPA action. 
Integration over the bosonic \textit{Gaussian} Action, in the presence of the current sources, we obtain 
\begin{equation}
    \ln Z_\text{RPA}[j] = i S_0[j] + \vec{A}^T_{[j]} iD \vec{A}_{[j]} - \frac{1}{2} \mathrm{Tr} \ln iD^{-1},
    \label{eq:RPA_action}
\end{equation}
where $S_0[j]$ is the mean-field contribution, given by 
\begin{gather}
    S_0[j] = \int dt ~ i \mathrm{Tr} \left[ \tilde{G}_0 \tilde{\Sigma}_0 [j] \right] + \frac{i}{2}
    \mathrm{Tr} \left[ \tilde{G}_0 \tilde{\Sigma}_0[j]  \tilde{G}_0 \tilde{\Sigma}_0[j] \right],
    \label{eq:RPA:Actions}
\end{gather}
The interaction contributions are given in terms of the Keldysh vector 
\begin{align}
    A^{\alpha}(t)_{[j]} &  = \notag \\
    & 2i\lambda \int dt_1 dt_2 \mathrm{Tr} \left[\tilde{G}_0(t_1,t) \gamma^{\alpha} \tilde{G}_0(t, t_2) \tilde{\Sigma}_{0}[j](t_2, t_1) \right]
    \label{eq:RPA:Aterm}
\end{align}
contracted with the bosonic propagator $D$.

\begin{figure*}[htbp]
    \includegraphics[width=\textwidth]{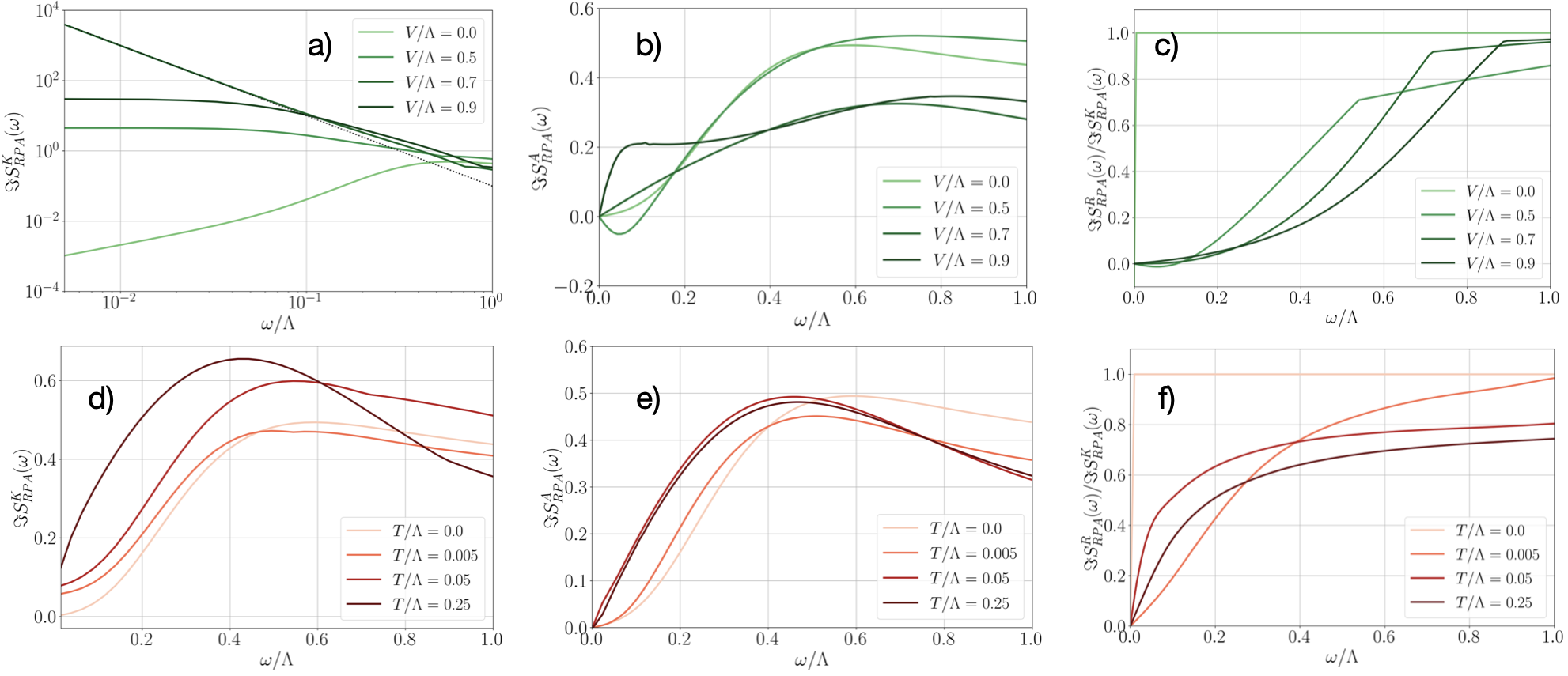}
    \caption{RPA Current Noise Correlations. Panels (a-c) show the $T=0$ case for several voltages ($V=0$, $V<V_c$, $V=V_c=0.7\Lambda$, $V>V_c$): (a) Imaginary part of the Keldysh component  $\Im \mathcal{S}^K_{\rm RPA}$ as a function of frequency $\omega$. (b) Imaginary part of the advanced response function  $\Im \mathcal{S}^A_{\rm RPA}$ as a function of frequency $\omega$. (c) The fluctuation-dissipation ratio $\Im \mathcal{S}^R_{\rm RPA}/\Im \mathcal{S}^K_{\rm RPA}$ as a function of frequency $\omega$.
    Panels (d-f) show the $V=0$(equilibrium) case for several temperatures:(d) Imaginary part of the Keldysh component  $\Im \mathcal{S}^K_{\rm RPA}$ as a function of frequency $\omega$. (e) Imaginary part of the advanced response function $\Im \mathcal{S}^A_{\rm RPA}$  as a function of frequency $\omega$. (f) The fluctuation-dissipation ratio  $\Im \mathcal{S}^R_{\rm RPA}/\Im \mathcal{S}^K_{\rm RPA}$  as a function of frequency $\omega$. }
    \label{fig:ShotRPA}
\end{figure*}

\subsection{Non-Interacting Contribution}
\label{subsec:MFNoise}

We first consider the non-interacting contribution of the current noise, obtained by varying the mean-field contribution of Eq.~(\ref{eq:RPA:Actions}):
\begin{equation}
\mathcal{S}^\text{K}_{0| L,L}(t,t') =  \frac{1}{2}\frac{\delta^2 S_0[j]}{\delta j^q_L(t) \delta j^q_L(t')} \bigg \lvert_{j=0}, 
\label{eq:S0K}
\end{equation}
where $j^c_L(t)$ ($j^q_L(t)$) is the \textit{classical} (\textit{quantum}) current source for the left lead at time $t$.
Similarly, the corresponding response functions are given by:
\begin{equation}
\mathcal{S}^\text{R}_{0| L,L}(t,t') =  \frac{1}{2}\frac{\delta^2 S_0[j]}{\delta j^{q}_L(t) \delta j^{c}_L(t')} \bigg \lvert_{j=0},   
\end{equation}

\begin{equation}
\mathcal{S}^\text{A}_{0| L,L}(t,t') =  \frac{1}{2}\frac{\delta^2 S_0[j]}{\delta j^{c}_L(t) \delta j^{q}_L(t')} \bigg \lvert_{j=0}.
\label{eq:S0A}
\end{equation}

The explicit expressions after performing the functional derivatives are given in the Appendix \ref{sec:Ap_CurrentNoise}. In the steady-state regime where time-translational invariance holds, we  perform a Fourier transform to the frequency domain
\begin{widetext}
\begin{gather}
    \mathcal{S}_0^\text{K}(\omega) =  i\int \frac{d\nu}{2\pi}\mathrm{Tr} \left[ \tilde{G}_0(\nu^-) \sigma_1 \Sigma_0(\nu^+) \sigma_1 \right]
    + \frac{i}{2}\int \frac{d\nu}{2\pi} \mathrm{Tr} \left[ \Sigma_0(\nu^+)\tilde{G}_0(\nu^+) \Sigma_0(\nu^+) \sigma_1 \tilde{G}_0(\nu^-) \sigma_1 \right] \notag \\
    + \frac{i}{2}\int \frac{d\nu}{2\pi} \left[
    \tilde{G}_0(\nu^+) \sigma_1 \Sigma_0(\nu^-) \tilde{G}_0(\nu^-) \Sigma_0(\nu^-) \sigma_1 \right] 
    - \frac{i}{2}\int \frac{d\nu}{2\pi} \mathrm{Tr} \left[ \Sigma_0(\nu^+)\tilde{G}_0(\nu^+)\sigma_1\Sigma_0(\nu^-) \tilde{G}_0(\nu^-)\sigma_1 \right] \notag \\    
    - \frac{i}{2}\int \frac{d\nu}{2\pi} \mathrm{Tr} \left[ \tilde{G}_0(\nu^+)\Sigma_0(\nu^+)\sigma_1\tilde{G}_0(\nu^-)\Sigma_0(\nu^-)\sigma_1 \right],
    \label{eq:CurrentNoise:NonIntKeldysh}
\end{gather}

\begin{gather}
    \mathcal{S}_0^\text{A}(\omega) =  i\int \frac{d\nu}{2\pi}\mathrm{Tr} \left[ \tilde{G}_0(\nu^-) \Sigma_0(\nu^+) \sigma_1 \right]
    + \frac{i}{2}\int \frac{d\nu}{2\pi} \mathrm{Tr} \left[ \Sigma_0(\nu^+)\tilde{G}_0(\nu^+) \Sigma_0(\nu^+) \tilde{G}_0(\nu^-) \sigma_1 \right] \notag \\
    + \frac{i}{2}\int \frac{d\nu}{2\pi} \left[
    \tilde{G}_0(\nu^+) \Sigma_0(\nu^-) \tilde{G}_0(\nu^-) \Sigma_0(\nu^-) \sigma_1 \right] 
    - \frac{i}{2}\int \frac{d\nu}{2\pi} \mathrm{Tr} \left[ \Sigma_0(\nu^+)\tilde{G}_0(\nu^+) \Sigma_0(\nu^-) \tilde{G}_0(\nu^-) \sigma_1 \right] \notag \\    
    - \frac{i}{2}\int \frac{d\nu}{2\pi} \mathrm{Tr} \left[ \tilde{G}_0(\nu^+)\Sigma_0(\nu^+)  \tilde{G}_0(\nu^-)\Sigma_0(\nu^-)\sigma_1 \right],
    \label{eq:CurrentNoise:NonIntAdvanced}
\end{gather}
\begin{gather}
    \mathcal{S}_0^\text{R}(\omega) = \left[ \mathcal{S}_0^\text{A}(\omega) \right]^{*}.
\end{gather}

\end{widetext}
where the frequency arguments are given by: $\nu^{\pm} = \nu \pm\omega/2$.

An expression reminiscent of  Eq.~(\ref{eq:CurrentNoise:NonIntKeldysh}) has been reported by Liu et al. in the context of a topological superconductor coupled to normal metal leads~\cite{Liu2018}.

\subsection{Interacting Contribution}
\label{subsec:IntNoise}

We now consider the interacting contribution to  the current noise of the RPA action in Eq.(\ref{eq:RPA_action}). 
Varying the interaction term in order to the sources $j$, we obtain: 
\begin{equation}
    \mathcal{S}^\text{K}_{int| L,L}(t,t') = \frac{1}{2}\frac{\delta^2  \vec{A}^T_{[j]} D \vec{A}_{[j]}  }{\delta j^q_L(t) \delta j^q_L(t')} \bigg \vert_{j=0} ,
\end{equation}
and
\begin{equation}
    \mathcal{S}^\text{A}_{int| L,L}(t,t') = \frac{1}{2}\frac{\delta^2  \vec{A}^T_{[j]} D \vec{A}_{[j]}  }{\delta j^c_L(t) \delta j^q_L(t')} \bigg \vert_{j=0} . 
\end{equation}
Explicitly in the steady-state, we get: 
\begin{align}
\mathcal{S}^\text{K}_{int} (\omega)   =  &
-2\lambda^2   \left[ D^\text{K}(\omega) T_+^{cl}(\omega) T_-^{cl}(\omega)  +  \right. \notag \\
& \left. D^\text{R}(\omega) T_+^{cl}(\omega) T_-^{q}(\omega) +  D^\text{A}(\omega) T_+^{q}(\omega) T_-^{cl}(\omega) \right], \label{eq:RPA_SK}\\
 \mathcal{S}^\text{A}_{int}(\omega)    = &  -2\lambda^2 D^\text{A} (\omega) T_+^{cl} (\omega) K_-^{q}(\omega), \label{eq:RPA_SA}
\end{align}
where we defined 
\begin{align}
    T^{\alpha}_{\pm} & = \mathcal{T}^{\alpha \pm}_1 - \mathcal{T}^{\alpha \pm}_2, \\ 
    K^{\alpha}_{\pm} & = \mathcal{K}^{\alpha \pm}_1 - \mathcal{K}^{\alpha \pm}_2,
\end{align}
with 
\begin{equation}
    \mathcal{T}_1^{\alpha \pm} (\omega)= \int \frac{d\nu}{\sqrt{2\pi}} \mathrm{Tr} \left( \left[ \tilde{G}_0 \gamma^{\alpha} \right] \left(\nu^{\pm} \right) \left[ \tilde{G}_0 \Sigma_0 \sigma_1 \right]\left(\nu^{\mp}  \right) \right),
    \label{eq:RPA_T1}
\end{equation}
\begin{equation}
    \mathcal{T}_2^{\alpha \pm} (\omega)= \int \frac{d\nu}{\sqrt{2\pi}} \mathrm{Tr} \left( \left[\Sigma_0 \tilde{G}_0 \gamma^{\alpha} \right] \left(\nu^{\pm} \right) \left[ \tilde{G}_0 \sigma_1 \right]\left(\nu^{\mp}\right) \right),
\end{equation}
\begin{equation}
    \mathcal{K}_1^{\alpha \pm} (\omega)= \int \frac{d\nu}{\sqrt{2\pi}} \mathrm{Tr} \left( \left[ \tilde{G}_0 \gamma^{\alpha} \right] \left(\nu^{\pm}\right) \left[ \tilde{G}_0 \Sigma_0 \right]\left(\nu^{\mp}  \right) \right),
\end{equation}
\begin{equation}
    \mathcal{K}_2^{\alpha \pm}(\omega) = \int \frac{d\nu}{\sqrt{2\pi}} \mathrm{Tr} \left( \left[\Sigma_0 \tilde{G}_0 \gamma^{\alpha} \right] \left(\nu^{\pm} \right) \left[ \tilde{G}_0  \right]\left(\nu^{\mp}  \right) \right),
    \label{eq:RPA_K2}
\end{equation}
where
$\nu^{\pm} = \nu \pm \omega/2$.
Although these  analytic expressions are quite involved, the numerical evaluation of the expressions above is straightforward requiring only  two  one-dimensional integrals in frequency space. 
In the following, we study these integrals numerically to obtain the current noise in different regimes. 

\subsection{Results  - Current Noise}

\begin{figure}[htbp]
    \includegraphics[width=0.4\textwidth]{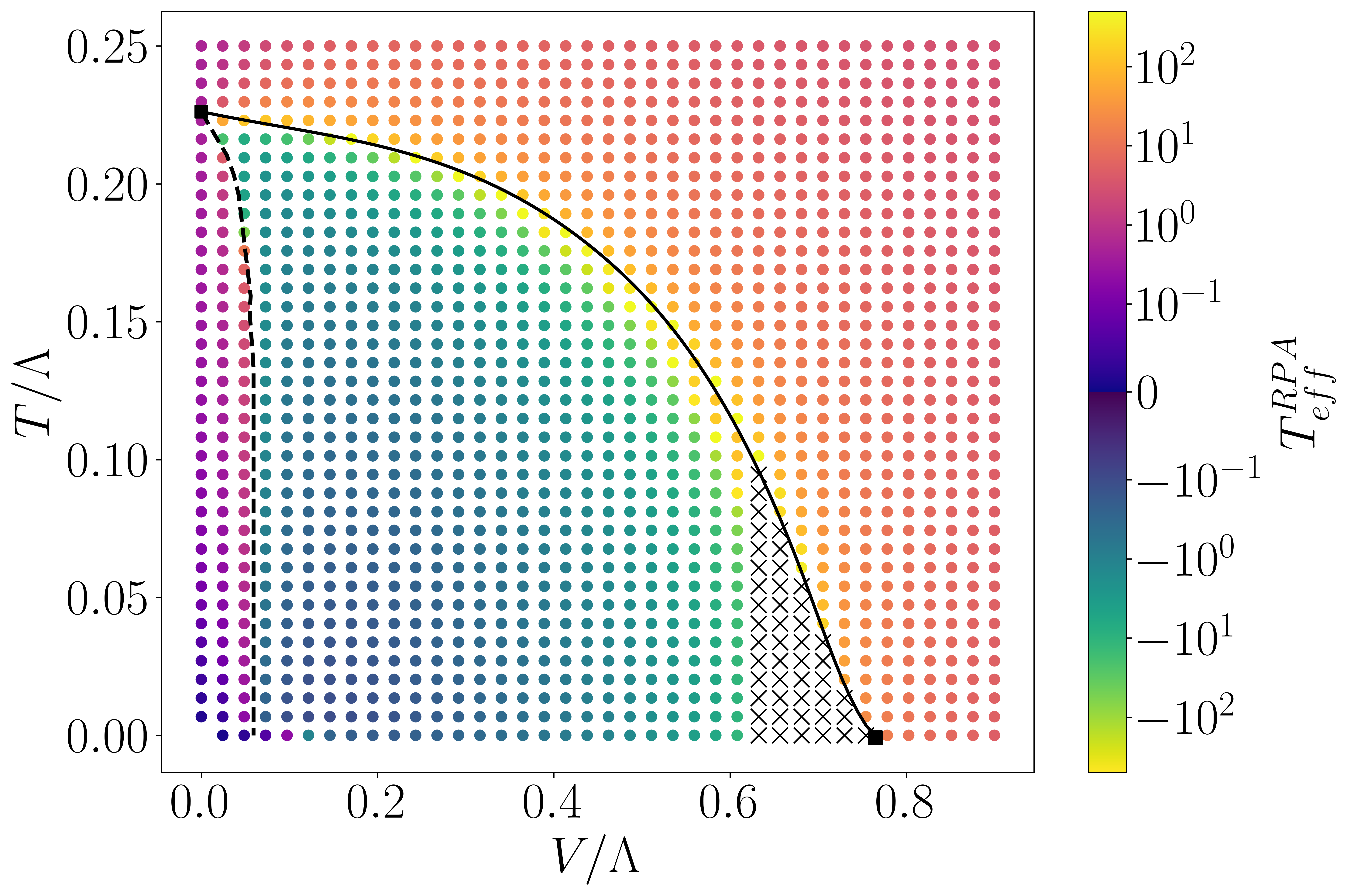}
    \caption{Zero-frequency effective temperature derived from current noise $T^{\mathcal{S}_{\rm RPA}}_{\rm eff}$ as a function of voltage $V$ and temperature $T$. The gray crosses indicate a region where results are omitted due to numerical precision issues.}
    \label{fig:ShotVT_2}
\end{figure}


Fig.~\ref{fig:ShotRPA} shows the correlation and response functions as a function of frequency at the RPA level. 

Deep in the disordered phase, at large temperatures or high voltages, the results resemble those for the non-interacting case. 
For comparison, the non-interacting case is given in Appendix \ref{sec:Non-int-Charge-noise}. 
For $T = 0$, $\Im S^K_\text{RPA}(\omega)$ exhibits a non-analyticity at $\omega = V$, below which the effective temperature (as defined by the current fluctuation-dissipation relation) is finite, and above which $T_\text{eff}(\omega) = 0$ is recovered. At finite temperature, this non-analyticity is smoothed into a crossover around $\omega \sim V$, with $T_\text{eff}(\omega \ll V) \gg T$ and $T_\text{eff}(\omega \gg V) \simeq T$.

In stark contrast to the non-interacting case, zero-frequency current fluctuations -- as measured by $\Im S^K_\text{RPA}(\omega)$ -- diverge (see Fig.\ref{fig:ShotRPA} - (a)), as we approach the transition. 
Surprisingly, the current response function -- as measured by $\Im S^A_\text{RPA}(\omega)$ -- becomes negative in an interval of positive frequencies (see Fig.\ref{fig:ShotRPA} - (b)). As a result, the fluctuation dissipation ratio becomes negative in that frequency interval (see Fig.\ref{fig:ShotRPA} - (c)). 
This corresponds to an effective temperature, $T^{\mathcal{S}_{\rm RPA}}_{\rm eff}$, that diverges at the transition and is negative in the non-equilibrium ordered phase. 

Fig.~\ref{fig:ShotVT_2} shows the effective temperature for $\omega=0$ as a function of $V$ and $T$. 
In the disordered phase $T_\text{eff}^{\mathcal{S}_{\rm RPA}}$ is positive and diverges at the transition. This is to be contrasted with $T_\text{eff}^{\chi_{\rm RPA}}$, which, although non-analytic, remains finite across the phase transition line. 
Within the ordered phase, $T_\text{eff}^{\mathcal{S}_{\rm RPA}}$ is negative for sufficiently large $V$. 
For $V=0$, $T_\text{eff}^{\mathcal{S}_{\rm RPA}}=T$ as required by the equilibrium fluctuation dissipation theorem (see Fig.\ref{fig:ShotRPA} - (d,e,f)).  
Interestingly, as $V$ increases, $T_\text{eff}^{\mathcal{S}_{\rm RPA}}$ decreases and becomes negative by crossing the  $T_\text{eff}^{\mathcal{S}_{\rm RPA}}=0$ line.  


Note that at sufficiently low temperature, and for voltages close to the transition, our numerical accuracy is not sufficient to resolve the zero-frequency fluctuation–dissipation relations. For this reason, we have marked these data points using a gray color code. Furthermore, fluctuations of the finite-$V$ zero-$T$ transition may affect this region. We, however, believe that the negative effective temperature is characteristic of the classically ordered phase.

\section{Conclusion}
\label{sec:Conclusion}

We investigate non-equilibrium phase transitions in a minimal quantum transport setup consisting of a particle–hole symmetric interacting quantum dot with a charging-energy term, tunnel-coupled to noninteracting leads. The fully connected character of the interaction renders the transition mean-field–like, even under finite bias voltage. We map out the non-equilibrium phase diagram as a function of temperature and voltage. Our main results concern the behavior of the charge susceptibility and the current noise across the phase diagram.

Interestingly, for the charge degrees of freedom, we find that an effective equilibrium description captures the charge correlations and response functions with an effective temperature that depends on the voltage and on the temperature of the leads, $T^\chi_\text{eff}(T,V)$. We note that not only the effective temperature, but also the finite-frequency fluctuation–dissipation ratios of a given observable are physically measurable quantities, see Appendix \ref{sec:Thermometry} for details.

The description in terms of an effective temperature implies that the order parameter susceptibility of the out-of-equilibrium transition is completely determined by its equilibrium counterpart. The Gaussian nature of the fixed point further implies that higher order correlators have to share this property. 
Using this description, we have established that the zero-frequency charge susceptibility diverges along the  phase transition line as $|V-V_c|^{-\gamma}$ or $|T-T_c|^{-\gamma}$, with $\gamma=2$, when crossing the transition along the voltage or temperature axes, respectively.

One may be concerned that the existence of an effective temperature is an artifact of the mean-field description. However, a related conclusion was reported in the context of a fully interacting quantum critical point, which is beyond a mean-field description \cite{Ribeiro_2013, Ribeiro2015}.

We also analyzed the current fluctuations and correlation functions away from equilibrium. Note that, current degrees of freedom are not determined solely by the order parameter and its fluctuations and have a non-local nature since they depend on the degrees of freedom of the leads. Thus, they are not solely determined by the critical fluctuations of the order parameter. 
Nevertheless, current fluctuations also diverge at the transition. 
Not surprisingly, the effective temperature measured by the current degrees of freedom behaves very differently from its charge counterpart. 
Most significantly, it becomes negative in the ordered side (sufficiently away from equilibrium). 
A natural interpretation of these results is that the temperature measuring device experiences a population inversion in the regions, resulting in a negative temperature. 

We note that a negative effective temperature is a consequence of the discrete nature of the order parameter, as it would become unstable for a continuous symmetry, where the transversal fluctuations diverge on the ordered side.

In summary, the charge susceptibility obeys an effective-temperature description in which its non-equilibrium divergence 
and scaling collapse mirror their equilibrium counterparts when expressed in terms of $T_{\mathrm{eff}}(T,V)$.
However, in sharp contrast with models where all observables share the same effective temperature,
we find that the current sector behaves in a fundamentally non-equilibrium manner: its fluctuation--dissipation 
ratio becomes negative in part of the ordered phase, implying a negative effective temperature associated with 
current fluctuations.  
This observable-dependent breakdown of $T_{\mathrm{eff}}$ highlights a key lesson emphasized in earlier studies:
while certain local degrees of freedom governed by the critical fixed point may admit an equilibrium-like 
organization, others—such as current operators that couple non-locally to both leads—can strongly violate
fluctuation--dissipation relations and retain genuinely non-equilibrium character \cite{MitraMillis2006,
KirchnerSi2009,RibeiroSiKirchner2013}.

Our findings demonstrate the value of current noise as a tool for detecting and probing critical fluctuations near quantum critical points, paving the way
for a better understanding of universality of critical systems out of equilibrium.

\section*{Acknowledgements}
S.K. acknowledges support by the National Science and Technology Council of Taiwan through Grant No.\ 112-2112-M-A49-MY4 and thanks NCHC for providing resources.
JA, and PR acknowledge support from FCT through the financing of the I\&D unit: Centro de Física e Engenharia de Materiais Avançados ref. UID/04540/2025 \footnote{https://doi.org/10.54499/UID/04540/2025}.

\appendix

\section{Markovian Regime}
\label{sec:Markovian}

In the large voltage regime, the dot dynamics becomes Markovian \cite{ribeiroNonMarkovianEffectsElectronic2015c}, and the self-energy $\Sigma_0$ becomes delta-correlated. As a result, the  dynamics is governed by a Lindblad equation, $\partial_t = \mathcal{L}[\rho]$, where $\rho$ is the density matrix and the Lindbladian operator, $\mathcal{L}$, is given by:
\begin{gather}
    \mathcal{L} [\rho] = -i [H_{dot}, \rho] + \Gamma \sum_{\alpha}  \left(  d^{\dagger}_{\alpha} \rho d_{\alpha} - \frac{1}{2} \{ d^{\dagger}_{\alpha} d_{\alpha}, \rho \}\right) + \notag \\
    +\Gamma \sum_{\alpha} \left(  d_{\alpha} \rho d_{\alpha}^{\dagger} - \frac{1}{2}\{ d_{\alpha} d^{\dagger}_{\alpha}, \rho \} \right).
\end{gather}
This treatment is asymptotically exact for $V\rightarrow \infty$, where the jump operators, $\hat{L}_{R\alpha} = d^{\dagger}_{\alpha}$ and $\hat{L}_{L\alpha} = d_{\alpha}$, correspond to the hoping from the fully filled lead to the dot and from the dot to the completely empty lead.  The Markovian limit significantly simplifies the analysis of the current and its fluctuations \cite{Landi2024},  however the limit $V\to\infty$ restricts the analysis to the disordered regime.
This helps explain why the current and current noise in the Markovian regime are the same as those obtained in  the non-interacting case:
\begin{equation}
\langle \hat{J}_L \rangle = \frac{\Gamma}{2},
\end{equation}
\begin{equation}
\mathcal{S}(\omega) = \Gamma - \frac{2\Gamma^3}{4\Gamma^2 + \omega^2}.
\end{equation}
These results agree with the those of Sec.\ref{subsec:MFNoise} in the large $V$ limit.

\section{Harmonic Oscillator Thermometry}
\label{sec:Thermometry}

Let us consider a single bosonic mode  at frequency $\Omega$ coupled to a large system. This could, {\itshape} i.e., be a local phonon coupled to the charge fluctuations on the dot of the main part.
The generating function is given by
\begin{align}
Z'& = \int DX\ Da\ e^{iS[X]+i\left\{ a^{\dagger}g_{0}^{-1}a-\eta\int_{C}dz\ X\left(z\right)\left[a\left(z\right)+a^{\dagger}\left(z\right)\right]\right\} }. 
\end{align}
Considering a small $\eta$, at the lowest order we can write
\begin{align}
Z'& = \int DX\ Da\ e^{iS[X]+i\left\{ a^{\dagger}g_{0}^{-1}a-\eta\int_{C}dz\ X\left(z\right)\left[a\left(z\right)+a^{\dagger}\left(z\right)\right]\right\} } \nonumber \\
  & \simeq   \int Da\ e^{i\left\{ a^{\dagger}g_{0}^{-1}a-\frac{\eta^{2}}{2}\left[a+a^{\dagger}\right]\chi\left[a+a^{\dagger}\right]\right\} } \\
  & = \int Da\ e^{i\int\frac{1}{2}A^{\dagger}\left(z\right) G(z,z') A\left(z'\right)}
\end{align}
where $\chi\left(z,z'\right) =-i \left\langle X\left(z\right)X\left(z'\right) \right\rangle $ is the contour-ordered correlator of the field the boson is coupled to. The last equality is obtained by writing the action in Nambu space $A = \{a,a^\dagger\}^T$, where the propagator given by
$$
G(z,z') = \left(\begin{array}{cc}
g_{0}^{-1}\left(z,z'\right)-\eta^{2}\chi\left(z,z'\right) & -\eta^{2}\chi\left(z,z'\right)\\
-\eta^{2}\chi\left(z,z'\right) & g_{0}^{-1}\left(z',z\right)-\eta^{2}\chi\left(z,z'\right)
\end{array}\right). 
$$
In the long-time limit when the system will have {\itshape thermalized} to its non-equilibrium steady-state limit, where time-translational invariance applies. 
In this limit and at the 
order of interest in $\eta$, the anomalous terms of the propagator vanish.
Therefore, it is sufficient to consider the diagonal part given by 
\begin{align}
g^{R}\left(\omega\right)&=\left[\omega-\Omega-\eta^{2}\chi^{R}\left(\omega\right)\right]^{-1}, \\ 
g^{K}\left(\omega\right)&= \frac{-\eta^{2}\chi^{K}\left(\omega\right)}{\left[\omega-\Omega-\eta^{2}\chi^{R}\left(\omega\right)\right]\left[\omega-\Omega+\eta^{2}\chi^{A}\left(\omega\right)\right]}.
\end{align}
For the Keldysh component we now use the identity $\lim_{\gamma\to0} \frac{\gamma/\pi}{x^{2}+\gamma^{2}}=\delta\left(x\right)$ with $\gamma = \eta^{2}\chi''\left(\omega\right)$. 
This yields 
$ g^{K}\left(\omega\right) = -2i\pi Q_{\text{FDR}}^{\chi}\left(\Omega\right)\delta\left(\omega-\Omega\right) $, where  $Q_{\text{FDR}}^{\chi}  = \chi^{K}/(\chi^{R}-\chi^{A})$ is the fluctuation dissipation ratio of $\chi$. 
The occupation of the bosonic level in the steady-state  
$$
\left\langle a^{\dagger}a \right\rangle =\frac{1}{2}\left[i\int\frac{d\omega}{2\pi}g^{K}\left(\omega\right)-1\right]=\frac{1}{2}\left[Q_{\text{FDR}}^{\chi}\left(\Omega\right)-1\right].
$$
It is worth noting that in equilibrium $Q_{\text{FDR}}^{\chi}\left(\Omega\right)=\tanh^{-1}\left(\frac{\beta\Omega}{2}\right)$, which simply states that the occupation is given in terms of the (equilibrium) Bose function at temperature $1/\beta$. 

Therefore, not only the effective temperature, but also the finite-frequency fluctuation–dissipation ratios of a given observable are physically measurable quantities. Indeed, this section shows they are directly related to the occupation of a harmonic oscillator weakly coupled to the corresponding degrees of freedom.

\begin{figure*}[t!]
    \includegraphics[width=\textwidth]{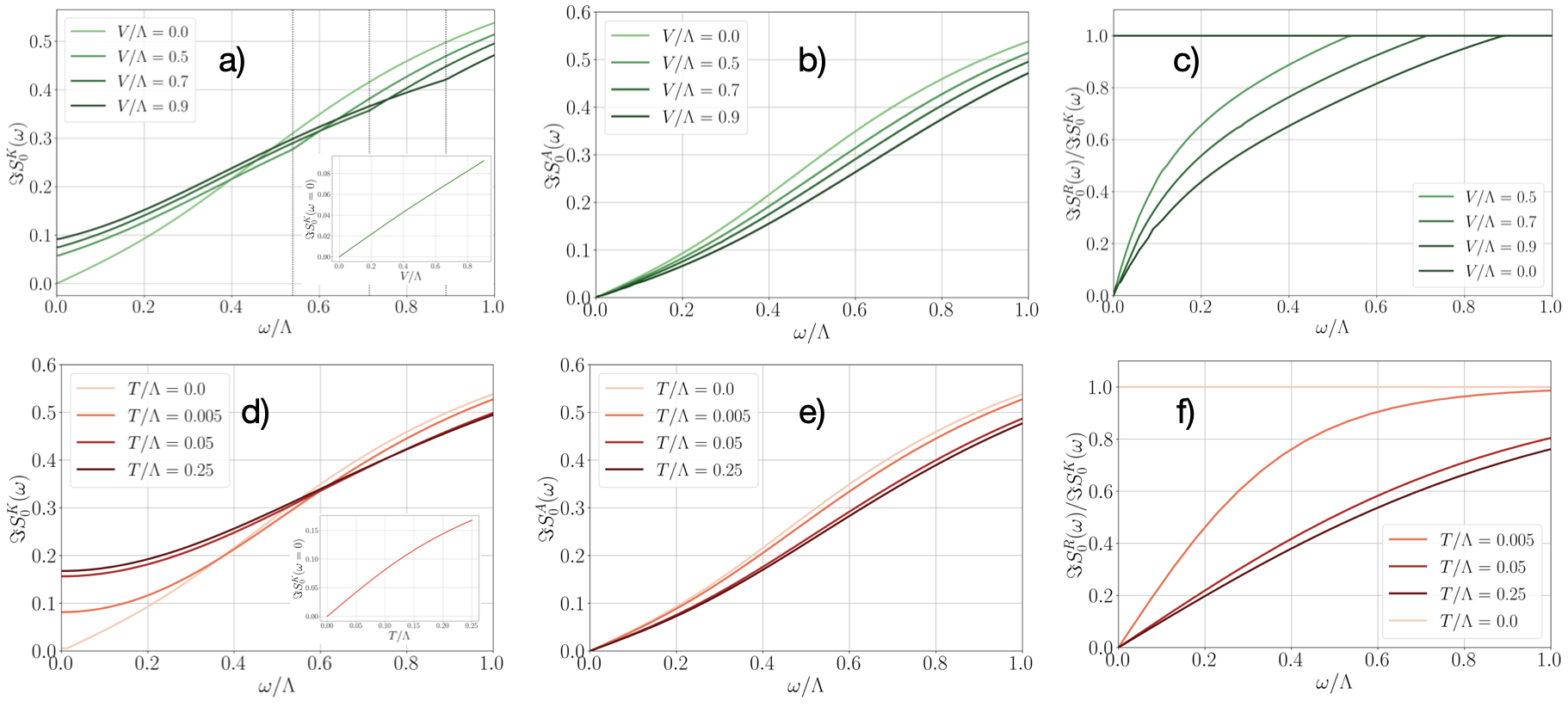}
    \caption{Non-interacting Current Noise Correlations. Panels (a-c) show the $T=0$ case for several voltages ($V=0$, $V<V_c$, $V=V_c$, $V>V_c$): (a) Imaginary part of the Keldysh component $\Im \mathcal{S}^K_0$ as a function of frequency $\omega$. The inset shows the zero-frequency value, $\Im \mathcal{S}^K_0(\omega=0)$, as a function of voltage $V$, highlighting its linear dependence. (b) Imaginary part of the advanced current noise function $\Im \mathcal{S}^A_0$ as a function of frequency $\omega$. (c) The fluctuation-dissipation ratio $\Im \mathcal{S}^R_0/\Im \mathcal{S}^K_0$ as a function of frequency $\omega$. Panels (d-f) show the $V=0$ (equilibrium) case for several temperatures: (d) Imaginary part of the Keldysh component $\Im \mathcal{S}^K_0$ as a function of frequency $\omega$. The inset shows the zero-frequency value,$\Im \mathcal{S}^K_0(\omega=0)$, as a function of temperature $T$. (e) Imaginary part of the advanced response function $\Im \mathcal{S}^A_0$ as a function of frequency $\omega$. (f) The fluctuation-dissipation ratio $\Im \mathcal{S}^R_0/\Im \mathcal{S}^K_0$ as a function of frequency $\omega$. }
    \label{fig:ShotNonInt}
\end{figure*}

\section{FDR for the RPA charge susceptibility}
\label{sec:FDR_RPA}

The fact that the RPA FDR and the bare one yield the same result can be simply obtained observing that
\begin{align*}
    \frac{\Im \chi^K_{RPA}}{\Im \chi^R_{RPA}} & =  
     \frac{ \Im \left[ \frac{\chi^K_0}{(1+4\lambda \chi^R_0)(1+4\lambda \chi^A_0)} \right]}{
    \Im \left[  \frac{\chi^R_0 (1+4\lambda \chi^A_0)}{(1+4\lambda \chi^R_0) (1+4\lambda \chi^A_0)} \right] } 
    = \frac{\Im \chi^K_0 }{\Im \chi^R_0},
\end{align*}
where we used $(\chi^R_0)^*  = \chi^A_0$ and thus:
\begin{align*}
    \Im \left[ (1+4\lambda \chi^A_0)(1+4\lambda \chi^R_0) \right]  = 0.
\end{align*}

\section{Explicit expressions of Current Noise correlations and response functions}
\label{sec:Ap_CurrentNoise}
\begin{widetext}
After performing the functional derivatives in Eqs.~(\ref{eq:S0K}, \ref{eq:S0A}) and reorganizing the expression, we obtain for the non-interacting contribution in the time domain
\begin{align}
\mathcal{S}^K_{0}(t,t') & =  i\mathrm{Tr} \left[ (G_0 \sigma_1)(t',t) (\Sigma_0 \sigma_1)(t,t')\right]  \notag 
+ \frac{i}{2} \mathrm{Tr} \left[ (\Sigma_0 G_0 \Sigma_0 \sigma_1) (t',t) (G_0 \sigma_1)(t,t') \right]  +  \frac{i}{2} \mathrm{Tr} \left[ (G_0 \sigma_1) (t',t) (\Sigma_0 G_0 \Sigma_0 \sigma_1)(t,t') \right] \notag  \\ 
& - \frac{i}{2} \mathrm{Tr} \left[ (\Sigma_0 G_0 \sigma_1)(t',t) (\Sigma_0 G_0 \sigma_1)(t',t) \right]   -\frac{i}{2} \mathrm{Tr} \left[ (G_0 \Sigma_0 \sigma_1)(t',t) (G_0 \Sigma_0 \sigma_1)(t,t') \right],
\end{align}
and
\begin{align}
\mathcal{S}_{0}^A(t,t') & = i\mathrm{Tr} \left[ G_0 (t,t') (\Sigma_0\sigma_1)(t',t) \right] \notag 
+ \frac{i}{2} \mathrm{Tr} \left[ (\Sigma_0 G_0 \Sigma_0)(t',t) (G_0 \sigma_1) (t,t') \right] \notag 
+ \frac{i}{2}\mathrm{Tr} \left[ (G_0) (t,t') (\Sigma_0 G_0 \Sigma_0 \sigma_1)(t',t)  \right] \notag \\
& - \frac{i}{2} \mathrm{Tr}\left[ (\Sigma_0 G_0)(t',t) (\Sigma_0 G_0 \sigma_1)(t,t')\right]   
- \frac{i}{2} \mathrm{Tr} \left[  G_0 \Sigma_0)(t',t) (G_0 \Sigma_0\sigma_1) (t,t')\right].
\end{align}

In the context of the expressions for interacting components of Current Noise, we begin with the following:
\begin{equation}
    \vec{A}^T_{[j]} D \vec{A}_{[j]} = -4\lambda^2 \int dt_1 dt_2  A_{[j]}^{\alpha}(t_1) D_{\alpha \beta} (t_1, t_2) A_{[j]}^{\alpha}(t_2).
\end{equation}

After calculating the $\delta A^{\alpha} (t_1) / \delta j^{\beta}_L (t_2)$ using Eq.~(\ref{eq:RPA:Aterm}), we consistently find a set of terms that exhibit the same temporal and causal structure, which is represented by:
\begin{equation}
    \mathcal{S}_{\mathrm{aux}}^{\alpha \beta}(t,t') =  -2\lambda^2 \int dt_1 dt_2 \,D_{\alpha \beta}(t_1, t_2) \, \mathrm{Tr} \left[ M^{\alpha}_1(t,t_1) M_2(t_1, t) \right] \,\mathrm{Tr} [ N^{\beta}_1(t',t_2) N_2(t_2, t')],
\end{equation}
where we should consider \( M_1^{\alpha} \) and \( N_1^{\alpha} \) as the first blocks of Green functions in Eqs. (\ref{eq:RPA_T1} - \ref{eq:RPA_K2}). They could correspond to either \( \left[\tilde{G}_0 \gamma^{\alpha}\right] \) or \( \left[\Sigma_0 \tilde{G}_0 \gamma^{\alpha}\right] \), depending on the specifics of \( \Sigma_0 \) and the Keldysh nature of the current source \( j \) (whether it is classical or quantum).
Following the same reasoning, \( M_2 \) and \( N_2 \) correspond to the second blocks in Eqs. (\ref{eq:RPA_T1} - \ref{eq:RPA_K2}) and may be represented by one of the following: \( \left[\tilde{G}_0 \Sigma_0 \sigma_1\right] \), \( \left[\tilde{G}_0 \sigma_1\right] \), \( \left[\tilde{G}_0 \Sigma_0\right] \), or \( \left[\tilde{G}_0\right] \).

In the steady-state and frequency domain, the expressions for $S^{\alpha \beta}_{\mathrm{aux}}$ always exhibit the same frequency structure, which is given by:
\begin{equation}
    \mathcal{S}_{\mathrm{aux}}^{\alpha \beta}(\omega) =  -2\lambda^2 D_{\alpha \beta}(\omega) \int \frac{d\nu_1}{\sqrt{2\pi}} \mathrm{Tr} \left[ M^{\alpha}_1(\nu_1^{+}) M_2(\nu_1^{-}) \right]  \int \frac{d\nu_2}{\sqrt{2\pi}}\mathrm{Tr}[  N^{\beta}_1(\nu_2^{+}) N_2(\nu_2^{-})].
\end{equation}

By applying this to $\mathcal{S}^{K}_{int}$ and $\mathcal{S}^{A}_{int}$, and after performing some algebraic manipulations, we derive the expression presented in Eqs. (\ref{eq:RPA_SK}, \ref{eq:RPA_SA}).

\end{widetext}

\section{Numerical Results for the  Current Noise in the Non-interacting Limit.}
\label{sec:Non-int-Charge-noise}

In this section we present the current fluctuations in the non-interacting limit ($\lambda=0$). 

Figure~\ref{fig:ShotNonInt} shows the correlation and response functions as a function of frequency.
At low temperatures and small voltages, the zero-frequency current noise, $\Im S_0(\omega=0)$, varies linearly with both $V$ and $T$ (see insets of panels (a) and (d)). While this behavior resembles the results obtained by Büttiker, the proportionality coefficients for shot noise, $\Im S_0(\omega=0, T=0) = s_V V$, and Nyquist noise, $\Im S_0(\omega=0, V=0) = s_T T$, differ from Büttiker’s expressions due to the fact that in our setup current fluctuations are measured near the quantum dot.
Nonetheless, the results are qualitatively similar. In particular, for $T = 0$, $\Im S_0(\omega=0)$ exhibits a non-analyticity at $\omega = V$, below which the effective temperature (as defined by the current fluctuation-dissipation relation) is finite, and above which $T_\text{eff}(\omega) = 0$ is recovered. At finite temperature, this non-analyticity is smoothed into a crossover around $\omega \sim V$, with $T_\text{eff}(\omega \ll V) \gg T$ and $T_\text{eff}(\omega \gg V) \simeq T$.
In the absence of interactions, we also have that the effective temperatures obtained from  the charge and current noise are qualitatively similar, differing at most by a factor of two for large voltages and low temperatures (see Fig.~\ref{fig:phy_vs_teff_ShotVT_non_interacting}).

\nocite{*}

\bibliography{main}

\end{document}